# Liquid Sensing Using WiFi Signals

YILI REN, Florida State University, USA
SHENG TAN, Trinity University, USA
LINGHAN ZHANG, Florida State University, USA
ZI WANG, Florida State University, USA
ZHI WANG, Florida State University, USA
JIE YANG, Florida State University, USA

The popularity of Internet-of-Things (IoT) has provided us with unprecedented opportunities to enable a variety of emerging services in a smart home environment. Among those services, sensing the liquid level in a container is critical to building many smart home and mobile healthcare applications that improve the quality of life. This paper presents LiquidSense, a liquid level sensing system that is low-cost, high accuracy, widely applicable to different daily liquids and containers, and can be easily integrated with existing smart home networks. LiquidSense uses existing home WiFi network and a low-cost transducer that attached to the container to sense the resonance of the container for liquid level detection. In particular, our system mounts a low-cost transducer on the surface of the container and emits a well-designed chirp signal to make the container resonant, which introduces subtle changes to the home WiFi signals. By analyzing the subtle phase changes of the WiFi signals, LiquidSense extracts the resonance frequency as a feature for liquid level detection. Our system constructs prediction models for both continuous and discrete predictions using curve fitting and SVM respectively. We evaluate LiquidSense in home environments with containers of three different materials and six types of liquids. Results show that LiquidSense achieves an overall accuracy of 97% for continuous prediction and an overall F-score of 0.968 for discrete predication. Results also show that our system has a large coverage in a home environment and works well under non-line-of-sight (NLOS) scenarios.

CCS Concepts: • **Human-centered computing** → **Ubiquitous and mobile computing systems and tools**.

Additional Key Words and Phrases: Internet-of-Things (IoT), Liquid Level Sesning, Smart Home, Resonance Frequency, WiFi

Authors' addresses: Yili Ren, Florida State University, 1017 Academic Way, Tallahassee, FL, 32306, USA, ren@cs.fsu.edu; Sheng Tan, Trinity University, One Trinity Place, San Antonio, TX, 78212, USA, stan@trinity.edu; Linghan Zhang, Florida State University, 1017 Academic Way, Tallahassee, FL, 32306, USA, lzhang@cs.fsu.edu; Zi Wang, Florida State University, 1017 Academic Way, Tallahassee, FL, 32306, USA, ziwang@cs.fsu.edu; Zhi Wang, Florida State University, 1017 Academic Way, Tallahassee, FL, 32306, USA, zwang@cs.fsu.edu; Jie Yang, Florida State University, 1017 Academic Way, Tallahassee, FL, 32306, USA, jie.yang@cs.fsu.edu.

1 INTRODUCTION

In recent years, the Internet of things (IoT) is becoming more and more intergraded into our daily life and is revolutionizing the way we live. By connecting everyday objects together, it provides a variety of emerging services to improve the quality of our life, especially in a smart home environment [8, 52, 61, 67]. Among those emerging services, sensing the liquid level in the containers has gained increasing attention [13, 24, 32, 33, 57] as it provides the information of when and how much the liquid content has been consumed each day. Such information is critical to building many smart homes and mobile healthcare applications [13]. For example, it helps us to estimate and track the calories ingested [36] or water consumed each day [49]. It can also help patients or care-taker to manage the intake and refill of the medicines [46] and to monitor the household inventory [37]. Integrating such information into a smart home environment could enable many novel applications, such as expiration notification and automatic reordering [1], which could greatly benefit our daily lives.

The challenge in liquid level sensing lies in finding low-cost and highly accurate solutions that are widely applicable to daily liquid containers and can be easily integrated with smart home networks. Existing commercial products are very expensive (e.g., the price is ranging from $50 to $100 for each container) as specialized sensors are used for each bottle [34, 50]. They are not scalable to a large number of daily liquid containers. Other industry solutions [11, 12, 16, 35, 57] are only suitable for large tanks in industry applications, which do not apply to the household settings. There are active research efforts in liquid level sensing with different technologies including the use of capacitive sensors [3, 5, 17, 29, 42, 54], pressure sensors [12, 35, 57], cameras [2, 8, 20–22] and acoustic sensors [4, 9, 25, 31]. However, the capacitive sensor-based approach [3, 5, 17, 29, 42, 54] requires the sensors to be immersed in the liquid to measure the capacitance of the container, which is both invasive and inconvenient as the immersed sensors may cause contamination, especially to the edible liquid. The camera-based approach [2, 8, 20–22], however, only works for transparent containers that filled with opaque liquid. Such an approach also suffers from performance degradation under the none-line-of-sight (NLOS) or poor lighting scenarios. Moreover, the system that utilizes pressure sensors [35] only works properly for containers with flat bottoms (i.e., where the pressure sensors attached to) and require careful calibration before each use. Finally, the acoustic-based approaches [4, 9, 25, 31] require specialized acoustic hardware to acquire the traveling time of ultrasound signals for liquid level detection. The specialized acoustic hardware, however, incurs a non-negligible cost. In addition, the above-mentioned approaches all require additional communication hardware for each container (e.g., WiFi or Bluetooth) to be integrated with existing smart home networks. Such a requirement incurs significant cost in addition to the dedicated sensors considering a large number of daily containers exist in a household setting.

In this paper, we introduce LiquidSense, a new approach for sensing the liquid level that is low-cost, high accuracy, widely applicable to different daily liquid containers, and can be easily integrated with existing home networks without additional cost. LiquidSense uses existing home WiFi networks and a low-cost transducer (i.e., only cost a few dollars) that attached to the container to sense the inherent vibration characteristic of the liquid in the container. Such a characteristic is closely associated with the liquid level and can be applied to a wide range of liquids as well as containers of different materials. Reusing existing home WiFi network for sensing liquid level allows LiquidSense to be integrated with existing smart home networks without additional communication hardware for each container, which provides tremendous cost-saving and can directly support a variety of smart home applications (e.g., automatic reordering). Though recent radio-frequency based sensing systems such as LiquID [10], RFIQ [18] and TagScan [56] can distinguish different types of liquids (e.g. water, milk, oil, and alcohol), they are not able to detect the liquid levels. LiveTag [15] can measure discrete water levels in a bottle by sensing the disturbed WiFi signals with a specialized metal tag. However, this system cannot provide continuous measurements and it is also constrained by the material of the bottle.

LiquidSense measures the resonance characteristic of the container that filled with liquid to infer the liquid level. The liquid level within a container is intimately linked to the container's resonance frequency, where the container vibrates with greater scale compared to other frequencies when it is exposed to an external forced vibration occurring at its natural frequency. Given a container filled with liquid, the mass of the liquid is the only factor that affects its natural frequency. We thus can infer the liquid level by measuring the resonance frequency of the container. In particular, LiquidSense leverages a low-cost transducer attached to the container's outer surface to generate external force vibration (i.e., chirp signal) to the container. The specially designed chirp signal allows the container to have high-frequency microvibration at its natural frequencies. Such microvibrations will introduce subtle changes to the signals in the home WiFi network. Through detecting the subtle changes in the WiFi signals, LiquidSense extracts the resonance frequency of the container for liquid level detection.

Accurately detecting the resonance frequency of the container is challenging when using the WiFi signals from the commodity WiFi devices. This is because the measurable changes in the WiFi signals caused by the high-frequency microvibration (about 2 to 3 millimeters vibrations) of the container are subtle. Existing commodity WiFi-based sensing systems can only discern human activities or locations at meter-level scale [38, 39, 58–60] or small motions (e.g., finger or mouth movements) at centimeter-level scale [28, 51, 55] with either the strength or doppler effect of the WiFi signals. They are not able to infer high-frequency microvibrations at about 2 to 3 millimeters scale. In this work, we designed a new sensing approach that utilizes the frequency analysis of the phase change at 5GHz WiFi channel to reveal the millimeter-level vibrations. First, the 5GHz channel has a shorter wavelength (i.e., 6cm), which increases discrimination power compare to the WiFi signals that commonly used at 2.4GHz channel. Moreover, with our proposed WiFi signal processing techniques, we are able to combine all the OFDM subcarriers to increase the signal-to-noise ratio and reliably detect the subtle change of the phase as small as 0.1 radians with phase-frequency analysis. Given the WiFi signals at 5GHz, this translates to the vibration detection at 0.95 millimeters scale, which provides sufficient resolution to capture the microvibration of the container for extracting the resonance frequency.

Given the extracted resonance frequency, LiquidSense applies novel matching algorithms to compare the measured resonance frequency against known profiles that identify the liquid level. In our system, we can perform either continuous liquid level or discrete liquid level prediction based on different frequency-liquid level profiles. For discrete prediction, we utilize Support Vector Machines (SVM) method to learn and predict liquid levels, whereas we explore the functional relationship between resonance frequency and liquid level (i.e., curvilinear regression) for continuous level prediction. We evaluate our system in typical home environments with three types of containers (i.e., metal, glass, and ceramic) and six types of liquids (i.e., water, coke, vegetable oil, milk, dishwashing liquid, and laundry detergent) under various scenarios (e.g., different distances and angles between WiFi devices and containers, both line-of-sight and non-line-of-sight conditions). The extensive experimental results show that our system achieves an overall accuracy of 97% for continuous prediction and an overall F-score of 0.968 for discrete predication. The contributions of our work are summarized as follows:

- We show that the resonance frequency of the container that filled with the liquid can be used to indicate the liquid level. This approach is non-invasive, low-cost, and is applicable to a wide range of liquids and different types of daily liquid containers.
- We build LiquidSense, which leverages the home WiFi networks to sense container microvibrations and capture the resonance frequency of the container for liquid level sensing. As WiFi signals are used for liquid sensing, it can be easily integrated with existing smart home networks without communication hardware for each container, which provides tremendous cost-saving in a smart home environment.
- We conduct extensive experiments under different experimental settings with six types of liquids and containers of three different materials. Results show that LiquidSense achieves an overall accuracy of 97% for continuous prediction and an overall F-score of 0.968 for discrete predication. LiquidSense also works well under different angles and large distances between the WiFi devices and the container as well as the NLOS scenarios.

## 2 RELATED WORK

In general, the approaches for liquid level detection or liquid sensing can be divided into six categories based on the sensing technologies they used: capacitive-based, pressure-based, camera-based, acoustic-based, vibration-based and RF-based approaches.

**Capacitive-based.** Capacitive sensors-based approach is widely used for sensing the liquid level in a container. However, most of these systems require a sensor to immerse in the liquid, which may contaminate the edible liquid. As the liquid is a dielectric material that affects the capacitance of the entire container, these systems measure the capacitance of the container with the immersed capacitive sensors to detect the liquid level [3, 5, 17, 29, 42, 54]. Recently, the systems proposed by Dietz *et al.* [11] and Geethamani *et al.* [16] can sense the liquid level with mounted or wrapped sensors outside of the container. However, those systems requires specifically designed hardware and circuit, which is not scalable to a large number of daily containers in a household setting. In addition, this approach requires additional communication hardware for each container to be integrated with existing home networks.

**Pressure-based.** This type of systems requires containers to have flat bottoms to attach the pressure sensors [35]. Moreover, to obtain accurate measurements, the pressure sensors need to be calibrated and reset before each use. As the pressure caused by the liquid is tied to the amount of liquid, these systems detect the liquid level by tracking the pressure change over time. Instead of installing the sensor at the bottom of the container, Wang *et al.* [57] present a system which is based on a pressure sensor connected to a floating pipe for water level detection. Likewise, Esmaili *et al.* proposed a pressure-based liquid level measuring instrument [12] which installs a sensor inside the container. These systems, however, all require direct contact with the liquid to be measured. Similar to the capacitive-based approach, pressure-based systems all require an additional communication unit for each container to communicate with a smart home network.

**Camera-based.** Camera-based methods require a transparent container filled with opaque liquids as well as good lighting conditions, which greatly limit the applicable scenarios. Chiu *et al.* proposed Playful Bottle [8], a camera-based liquid level detection system which can use a smartphone camera to capture a transparent bottle with pattern bars. Hannan *et al.* [21] proposed a waste bin level detection systems leveraging a camera as an image capturing device and a Multi-Layer Perceptron (MLP) classifier as an inference algorithm. Other systems [2, 20] mount a camera above the waste bin and can distinguish states of waste bins using K-nearest neighbor (KNN) classifiers. Also, Jiang *et al.* [22] proposed a system that can measure the level of food in a container using deep convolutional neural networks (CNNs) trained images captured by an off-the-shelf camera.

**Acoustic-based.** Most of the acoustic-based systems measure the Time-of-Flight (ToF) of the acoustic signal travel through the container for liquid level sensing [4, 9, 25, 31]. However, these systems require specialized hardware to measure the ToF, which incurs a non-negligible cost. Besides the ToF based approach, Li *et al.* proposed a liquid level detection system [26] based on the ultrasonic scattering theory and the boundary-layer theory. Fan and Truong proposed SoQr [13], an ultrasonic based content level sensing system to detect the content level of a common household by using Mel-Frequency Cepstral Coefficients (MFCC) and SVM classifiers. Similar to previous solutions, those systems also require additional communication module to work in the smart home environment.

**Vibration-based.** Ryu *et al.* developed a non-contact type polyvinylidene fluoride (PVDF)-based liquid volume sensor [45]. When a container receives an excitation and generates a vibration, the sensor that is mounted on the wall of the container can measure the resonance frequency of vibration, which indicates the liquid level. However, a PVDF-based sensor is a specialized sensor, which increases the system cost. Zhao *et al.* proposed VibeBin [67], a vibration-based waste bin content level detection system. VibeBin attaches a vibrating motor on the waste bin and uses a vibration sensor to sense the waste bin vibrations to infer the content level. However, this system can only infer a few discrete fill-levels but cannot provide a continuous level prediction.

**RF-based.** Radio-frequency (RF) signals have been used to sense the liquid levels as well as the type of liquids. For example, leveraging the remote-sensing capability of millimeter waves, Nakagawa *et al.* proposed a contactless method [33] to measure the liquid level. But this method requires a millimeter Doppler sensor and a piezoelectric vibrator, which is not available to household settings. Mukherjee's system [32] sends radio waves from the bottom of the container to analyze the phase angle change of the reflected waves. However, this system requires to place two pieces of electrodes around the target container. LiveTag [15] is able to detect discrete water levels in a bottle using a specialized metal tag which can disturb WiFi signals. However, this system cannot monitor continuous water level changes and it only works on limited types of container materials. For example, a metal container would block WiFi signals, thus invalidating this system. Recently, several systems are proposed to identify the types of liquid types (e.g. water, milk, oil, and alcohol) instead of the liquid level [10, 18, 41, 56, 66]. For example, LiquID [10] can identify liquids by analyzing the time delay of Ultra-Wideband (UWB) signals which pass through the liquids. RFIQ [18] can identify liquids using RF coupling with Ultra-High Frequency (UHF) signals. TagScan [56] is an RFID-based system which can leverage the phase and RSS changes of RF signals to identify different liquids. Nutrilyzer [41] is a photoacoustic sensing system for detecting the nutrients and measuring adulterants in liquid food. CapCam [66] can measure liquid surface tension by using a smartphone and detect alcohol concentration and water contamination.

## 3 PRELIMINARIES
### 3.1 Resonance Properties of Objects

Our system exploits the resonance characteristic of the container that filled with liquid to infer the liquid level. Resonance refers to the phenomenon when a physical system vibrates at specific frequencies where an externally applied force is equal or approximate to the natural frequencies of the system [63]. Those frequencies are called resonance frequencies at which the system vibrates at a relative maximum amplitude compare to other frequencies. It has been shown in previous work [23, 53] that when a container is filled with liquid, both the frequency and amplitude of the resonance will vary depending on the liquid level. This is due to the mass of liquid in the container is the only factor that affects the natural frequency. The phenomenon of resonance thus can be used as an indicator of liquid level inside the container.

We attach a low-cost transduce on the outer wall of the container to emit a well-designed chirp signal to make container vibrates at resonance frequencies. Based on our experiments and existing research [23, 43], there are multiple resonance frequencies associated with an object. Among those frequencies, the first one is the strongest and the most stable. Figure 1 illustrates the frequency response of the first and the second resonance frequencies induced by a contact vibration sensor that attached to a metal container. It is obvious that the power level of the first resonance frequency is much stronger than that of the second resonance frequency. We thus sense and extract the first resonance frequency in our system to derive the liquid level inside a container.

There are two types of measurements can be used to derive liquid level based on resonance characteristics of the container: 1) the power of the resonance frequencies; 2) the value of the specific resonance frequencies. Figure 2 shows both the power level and the frequency value when a container is filled with water at different levels and vibrates at its first resonance frequency. As the increasing of liquid volume leads to the increase of container mass, both the power and the value of the resonance frequency should decreases. However, it is very hard to observe such a linear relationship between the measured power and the liquid level due to the low-cost transducer cannot provide very stable output power. For example, the power level of vibration with 1400ml liquid is larger than 600ml and the power level of vibration with 1000ml and 1200ml are very similar. This means the power of the resonance frequencies is not a very accurate liquid level indicator due to the use of a low-cost transducer. Instead, the value of the first resonance frequency is strongly correlated with the liquid level, as shown in Figure 2. In particular, the frequency value is negatively correlated with the liquid level. Therefore, we leverage the value of the resonance frequency instead of the power to infer the liquid level.

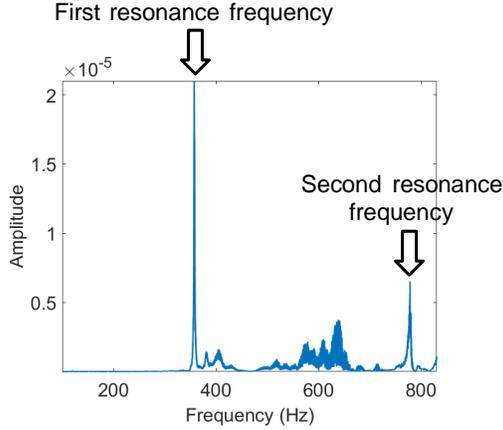

Fig. 1. Frequency response between 100Hz and 850Hz of a vibrating metal container measured with a contact vibration sensor.

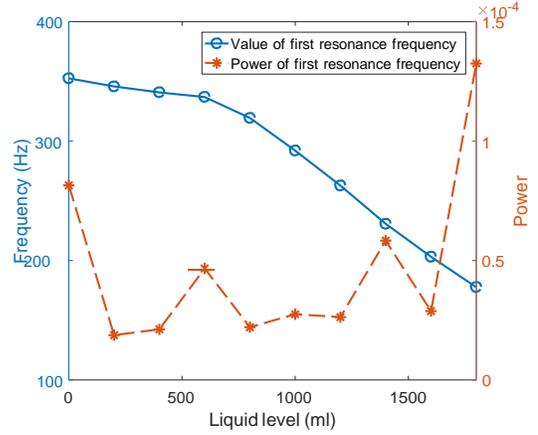

Fig. 2. Power and value of the first resonance frequency at different water levels of a vibrating metal container.

### 3.2 WiFi-Based Sensing Technique

The proliferation of wireless devices and networks provide us the opportunity to use the WiFi signals to capture the motions in physical space at various scales. We thus reuse the WiFi signals in a home network to sense the microvibrations of the container. By leveraging the multiple antennas readily available on many WiFi devices and the WiFi signals at 5 GHz, we can perform accurate sensing of the resonance frequency. As the WiFi signals can pass through a wall, using WiFi signals to sense container vibration can achieve non-contact sensing and work well under non-line-of-sight (NLOS) scenarios across rooms. Moreover, reusing WiFi for liquid level sensing presents tremendous cost-saving as the liquid level information can be directly fed into the home network for building smart home applications.

To detect the microvibrations of the container, we exploit the Channel State Information (CSI), which describe how the WiFi signals are changed after multipath propagation in the physical space. The CSI is tracked by off-the-shelf WiFi network interface cards (NICs) in 802.11 n/ac WiFi networks with multiple transmitting ($Tx$) and receiving ($Rx$) antennas. Such information can also be viewed as a sampled version of the channel frequency response (CFR). On the standard 20/40MHz WiFi channel between one transmission antenna pair, the CSI measurement includes both the amplitude and phase information for each of the 56/128 orthogonal frequency division multiplexing (OFDM) subcarriers. In this work, we utilize three receiving antennas on a standard 20MHz channel at 5GHz band. By leveraging multiple antennas, we able to extract more information and mitigate the CSI errors caused by commodity hardware limitation.

In particular, we denote $N_{Tx}$ as the number of transmit antennas, $N_{Rx}$ represents the number of receive antennas and $S$ indicates the number of OFDM subcarriers. For each subcarrier, the WiFi channel can be represented by $\mathbf{y} = H \mathbf{x} + \mathbf{n}$, where $\mathbf{y}$ is the received signal, $\mathbf{x}$ is the transmitted signal, $H$ is the CSI matrix with dimensions $N_{Rx} \times N_{Tx}$, which represents a complex valued matrix of channel frequency response, and $\mathbf{n}$ is the noise vector. As CSI is measured on 30 selected OFDM subcarriers for Intel 5300 WiFi NIC, each CSI measurement contains 30 matrices with dimensions $N_{Rx} \times N_{Tx}$. For example, the following matrix represents a nominal CSI measurement at time $t$ reported by Intel 5300 WiFi NIC for $N_{Tx} = 1$, $N_{Rx} = 3$ and $S = 30$ subcarriers,

$$\text{CSI measurement} = \begin{bmatrix} H_{1,1} & H_{1,2} & \dots & H_{1,30} \\ H_{2,1} & H_{2,2} & \dots & H_{2,30} \\ H_{3,1} & H_{3,2} & \dots & H_{3,30} \end{bmatrix} = \begin{bmatrix} H_1 \\ H_2 \\ H_3 \end{bmatrix}, \qquad (1)$$

where $H_{m,n}$ is the complex valued CFR for the $m^{\text{th}}$ receiving antenna and the $n^{\text{th}}$ subcarrier ($m = 1, 2, 3$, $n = 1, 2, ..., 30$). Also, $H_m$ represents the CFR values on the $m^{\text{th}}$ receiving antenna. As we known, WiFi signals are affected by the multipath propagation and the motions in the physical environments. Thus, tracking the changes in the CSI measurements enables us to derive the characteristics of the motions. If a WiFi signal arrives at the receiver through $N$ different paths, one entry in the matrice of CSI measurement can be written as:

$$H_{m,n}(f_n, t) = \sum_{k=1}^{N} a_k(f_n, t) e^{-j2\pi f_n \tau_k(t)}, \qquad (2)$$

where $f_n$ is the frequency on the $n^{\text{th}}$ subcarrier, $t$ is the time, $a_k(f_n, t)$ is the attenuation factor of the $k^{\text{th}}$ path. $e^{-j2\pi f_n \tau_k(t)}$ is the phase shift of the $k^{\text{th}}$ path which is caused by the propagation delay of $\tau_k(t)$. CSI amplitude and phase are impacted by multipath effects including amplitude attenuation and phase shift. This provides the theoretical basis for WiFi sensing.

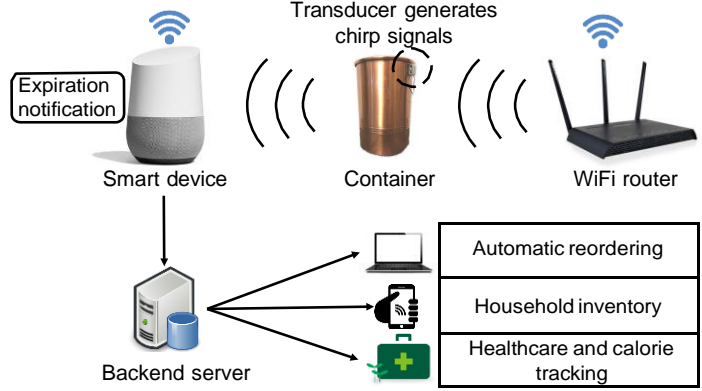

Fig. 3. System overview.

## 4 SYSTEM DESIGN

In this section, we discuss in detailed the system overview, the excitation signal design, the methods used to sense the microvibrations of the container, and the liquid level prediction models.

### 4.1 System Overview

The basic idea of our system is to use the home WiFi networks to sense the microvibrations of the container that stimulated by the attached low-cost transducer for liquid level detection. Recent advances in wireless technology have greatly expanded WiFi usage from providing laptop connectivity to connecting smart devices such as Google Home, Amazon Echo, surveillance cameras, refrigerators and smart TVs to a home network and the Internet. This provides ubiquitous coverage of WiFi links inside homes. Moreover, current WiFi radios (e.g., 802.11 n/ac) provide fine-grained channel measurements that allow us to track minute changes in the signal reflections due to the physical motions. This inspires us to reuse existing home WiFi links, such as the link between the WiFi router and the Google Home or Amazon Echo, to sense the microvibrations of the container for liquid level detection.

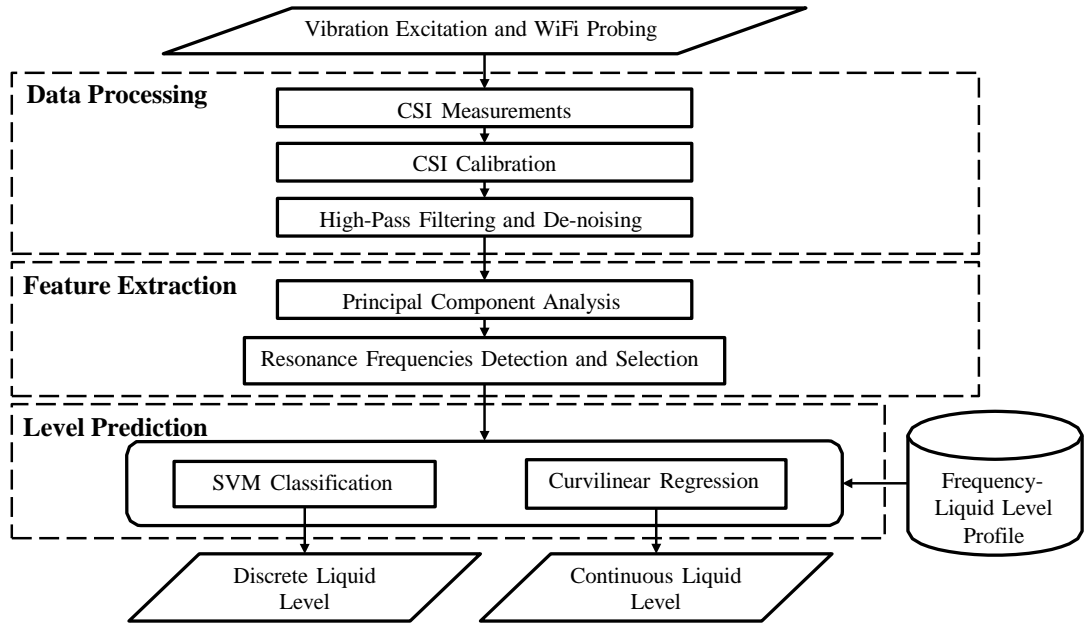

Fig. 4. System flow.

Figure 3 shows an overview of our system. We first attached a small transducer on the outer surface of the container, which is both low-cost and easy to install. The transducer then emits the excitation signal, which generates external force vibration to the container. The well-designed excitation signal allows the container to have high-frequency microvibration at its natural frequencies. As the surface of the container reflects the WiFi signals, such microvibrations will introduce subtle changes to the signal propagation, which in terms affects the received WiFi signals at the receiver. By analyzing the subtle changes in the WiFi signals, we can infer the microvibration of the container and extract the resonance frequency for liquid level detection.

Reusing WiFi network for liquid level sensing has a unique advantage as the liquid level information can be directly fed into the home network for building smart home applications. For example, the Google Home or Amazon Echo can directly utilize the liquid level information for automatic reordering or expiration notification services. Or the WiFi devices that sensed the liquid level information can send it to a smart home backend server to enable other novel smart home and mobile healthcare applications, such as calorie tracking, household inventory monitoring. Therefore, our system presents tremendous cost-saving in a household setting as no need for each daily container to have an additional communication unit to be connected with smart home networks.

The system flow is illustrated in Figure 4. It includes four major components: *Vibration Excitation and WiFi Probing*, *Data Processing*, *Feature Extraction* and *Level Prediction*. First, we use a carefully designed excitation signal to induce vibration of a container. We adopt the swept sine chirp with a bidirectional sweeping manner for the excitation signal. Our system then utilizes commodity WiFi to probe the target container and extract CSI measurements at the receiver end.

The received CSI measurements go through *Data Processing* component to remove the noises as well as the environmental interferences. Our system first takes the time-series CSI measurements extracted from one pair of commodity WiFi devices as input. The extracted CSI is calibrated to remove phase offsets which are caused by the unsynchronized clocks of the transmission pair. To handle environmental noises, our system uses a high-pass filter to remove lower frequency noises introduced by human activity and use spectral subtraction to further filter out other noises.

Next, our system achieves *Feature Extraction* by utilizing Principal Component Analysis (PCA) as well as resonance frequencies detection and selection. Since the vibration of the container even at resonance frequencies is still a microvibration (i.e., at millimeter-level), it is difficult to extract any useful information using only denoised CSI measurements. Thus, we adopt PCA to further enhance the signal components associated with the subtle vibration motion. After enhancement, our system strategically extracts only the first resonance frequency which is the strongest and most stable among all as the feature for later liquid level detection.

At last, our system conducts *Level Prediction* for liquids in both discrete and continuous manners. In particular, we use Support Vector Machine (SVM) to achieve the classification for discrete prediction. On the other hand, we leverage the curve fitting to construct the model for the relationship between resonance frequency and liquid level. Then such model is utilized for continuous prediction.

## 4.2 Excitation Signal Design

In this section, we discuss which type of excitation is suitable for our system and how to design such a signal.

**Type of Excitation.** Commonly used excitation signals can be divided into three categories: random excitation, impulse excitation, and sinusoidal excitation [14]. For the first category, the signal is random with Gaussian distribution and contains all frequencies within the frequency range. However, it is difficult to control the force of a random excitation. Impulse excitation is widely used due to its simplicity. But it suffers from shortcomings such as low signal-to-noise ratio (SNR) and poor repeatability. Alternatively, the sinusoidal excitation is an ideal alternative due to better control over the force and high repeatability. In this work, we adopt linear sweeping of sinusoids which is known as a swept sine chirp. The waveform of a swept sine chirp $x(t)$ is defined as:

$$x(t) = P \sin\left[2\pi\left(f_S t + \frac{1}{2}\varepsilon t^2\right)\right], \tag{3}$$

where $P$ is the amplitude of force and $f(t)$ is the frequency of the swept sine. Since the excitation frequency is a linear function of time, the sweep rate $\varepsilon$ is constant and can be expressed as

$$\varepsilon = \frac{f_E - f_S}{\tau}, \tag{4}$$

where $f_S$ and $f_E$ are the start and end frequencies respectively and $\tau$ is the excitation duration. Note that, the sweep rate $\varepsilon$ can be positive in the case of frequency increase, or negative in the case of frequency decrease.

**Chirp Design.** In designing the swept sine chirp signal, we primarily consider three factors: impact on system accuracy, frequency band and sweep rate of chirp. Based on previous research [30], the maximum amplitude is achieved only after the excitation frequency $f(t)$ equals to the resonance frequency $f_R$ and last for a certain duration of time. Since it is difficult to acquire the time duration, this could lead to resonance frequency estimation error. Thus we adopt a bidirectional excitation method [30] to mitigate such error. Assume excitation has the positive sweep rate $\varepsilon^+$, the frequency that has maximum amplitude is $f(t) = f^{\varepsilon^+}$. When the excitation has the negative sweep rate $\varepsilon^-$, the frequency that has maximum amplitude is $f(t) = f^{\varepsilon^-}$. Let $\Delta f^{\varepsilon^+} = f^{\varepsilon^+} - f_R$ and $\Delta f^{\varepsilon^-} = f_R - f^{\varepsilon^-}$. Since we use the excitation with $\varepsilon^+ = |\varepsilon^-|$, the equation $\Delta f^{\varepsilon^+} = \Delta f^{\varepsilon^-}$ is established. It is easy to extract the frequencies $f^{\varepsilon^+}$ and $f^{\varepsilon^-}$ from the signal portions with biggest amplitudes obtained for excitation with positive and negative sweep rates, respectively. Based on the symmetry of the frequency shift $\Delta f^{\varepsilon^+}$ and $\Delta f^{\varepsilon^-}$ to the resonance frequency, the resonance frequency can be expressed as:

$$f_R = \frac{f^{\varepsilon^+} + f^{\varepsilon^-}}{2}. \qquad (5)$$

According to existing work [23, 53] and our observations, the resonance frequency of commonly used household containers range from 140Hz to 900Hz. Thus, we choose 0Hz as the starting frequency and 1000Hz as the ending frequency. Furthermore, we choose 2000 pkts/s as the packet rate for the WiFi probing based on Nyquist–Shannon sampling theorem [47] to cover the frequency range of 1000Hz.

The sweep rate of the chirp impacts both the detection time and the accuracy of our system. Since a slower sweep rate will consume more energy of the transducer and lead to a longer system delay. Meanwhile, a fast sweep rate will result in performance degradation of the system. Thus, we empirically choose a sweep rate of $|\varepsilon| = 66.67$Hz/s. As discussed before, the accurate resonance frequency can be calculated by averaging the frequency values of both positive and negative sweep rates. Therefore, we set $f_S = 0$Hz, $f_E = 1000$Hz and $\varepsilon = 66.67$Hz/s for the positive sweep while $f_S = 1000$Hz, $f_E = 0$Hz and $\varepsilon = -66.67$Hz/s for the negative sweep.

### 4.3 Wireless Vibrometry

The basic idea of wireless vibrometry [62] can be viewed as modulation/demoluation of the object vibration using the wireless signals' amplitude and phase information. It is similar to the process used by laser radar (LADAR) [6] where a laser beam is emitted toward the vibrating object and the vibration can be inferred based on the reflected beams. Differently, the WiFi signal is less directive and suffer from reflection/diffusive loss. On the other hand, WiFi signal is more sensitive to the multipath overlapping patterns. Therefore, the wireless vibrometry can be viewed as a process where vibrations modulate the wireless signal in the multipath effect. Specifically, the amplitude and phase of the WiFi signals affected by the vibration displacement of the object will contain the vibration frequency information of such an object.

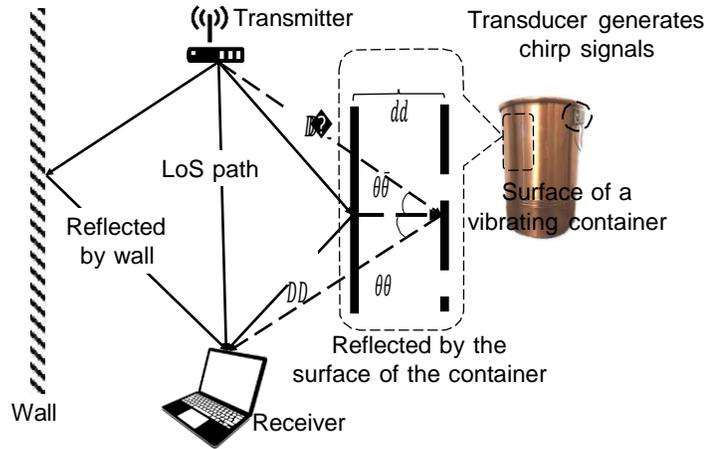

Fig. 5. Multipath propagations of WiFi signals when a container is vibrating.

**Vibration Displacement.** In this paper, vibration displacement is defined as the surface displacement of a vibrating container which is at millimeter-level. We assume that the transducer is vibrating at a swept sine chirp with a frequency $\omega = f_S + \frac{1}{2}\varepsilon t$ at time $t$. The outer surface of container vibrates at the same frequency $\omega$ at time $t$. The vibration displacement at time $t$ can be written as:

$$d(t) = A(\omega)\sin(\omega t + \varphi), \tag{6}$$

where $A(\omega)$ is the vibration magnitude and $\varphi$ is the initial phase of the chirp signal. Because resonance frequencies affect the vibration magnitude, vibration magnitude is determined by $\omega$. For simplicity, we omit the time $t$ of functions in later notations and $d(t)$ can be denoted as $d$.

**Impact of Vibration on Wireless Signal.** Figure 5 illustrates a typical experimental setup of sensing liquid level using a pair of WiFi devices. As demonstrated in the figure when the surface of the container is vibrating, it will affect the reflected signal. For simplicity purposes, we only consider the scenario where wireless signal is propagating on a single subcarrier $f$ and the $k^{\text{th}}$ path which has a length $D$. At time $t$, the propagating signal can be represented by a sinusoid $S(f,t) = a_k(f,t)e^{-j2\pi ft}$, where $a_k(f,t)$ is the attenuation of the path $k$ and $e^{-j2\pi ft}$ is the phase. Let $X(f)$ represent the signal when it was initially transmitted at time $t = 0$. $X(f)$ can be written as $X(f) = S(f,0) = a_k(f,0)e^{-j2\pi f \times 0}$. Let $Y(f)$ be the received signal when it arrives at $t = T$ (propagation delay) after traveling a distance of $D$. Hence, $Y(f)$ can be written as $Y(f) = S(f,T) = a_k(f,T)e^{-j2\pi fT}$. It is well known that the amplitude of a wireless signal is inversely proportional to the square of the distance it travels. Hence, we have $a_k(f,T) = p\frac{a_k(f,0)}{D^2}$, where $p$ is the proportionality constant. $e^{-j2\pi fT}$ is the phase shift of the $k^{\text{th}}$ path which is caused by the propagation delay. Let $\lambda$ be the wavelength, we have $\lambda = c/f$ and $T$ can be written as $T = D/f\lambda$. Then the phase shift of the $k^{\text{th}}$ path is $e^{-j2\pi fT} = e^{-j2\pi D/\lambda}$. Therefore, we have $Y(f) = p\frac{a_k(f,0)}{D^2}e^{-j2\pi D/\lambda}X(f)$. Since $Y(f) = H(f)X(f)$, the CFR can be written as:

$$H(f) = \frac{p}{D^2}e^{-j2\pi\frac{D}{\lambda}}. \tag{7}$$

During the sensing process, the $k^{\text{th}}$ path is disturbed by the vibration of the container. As shown in Figure 5, the vibration displacement of the surface of the container is $d$ and the length of the $k^{\text{th}}$ path changes from $D$ to $\bar{D}$. We can calculate $\bar{D}$ approximately: $\bar{D} \approx D + d(\cos\theta + \cos\bar{\theta})$, where $\theta$ and $\bar{\theta}$ are indicated in Figure 5. Replacing $D$ in Equation (7) with $\bar{D}$, we get

$$H(f) = \frac{p}{(D + d(\cos\theta + \cos\bar{\theta}))^2}e^{-j2\pi\frac{D+d(\cos\theta+\cos\bar{\theta})}{\lambda}}. \tag{8}$$

Previous equations only consider the signal propagation with single path. However, in reality, due to multipath affect, there are multiple different paths when wireless signals propagate from the transmitter to the receiver. Hence, the frequency response of the received signal is the superimposition of multiple paths. As shown in Figure 5, the multiple paths that the signal traverses can be divided into two categories: static component and dynamic component. The static component includes both LoS paths and the paths reflected from the surrounding wall. Dynamic component includes the paths disturbed by the vibrating container. Thus the CFR of received signal can be written as:

$$H(f) = H_s(f) + H_d(f) = H_s(f) + \sum_{k \in P_d}\frac{p}{(D_k + d(\cos\theta_k + \cos\bar{\theta}_k))^2}e^{-j2\pi\frac{D_k+d(\cos\theta_k+\cos\bar{\theta}_k)}{\lambda}}, \tag{9}$$

where $f$ is the frequency of the subcarrier, $P_d$ is the set of dynamic paths, $D_k$ is the total length of the $k^{\text{th}}$ path and $\theta_k$ and $\bar{\theta}_k$ are the corresponding angles. $H_s(f)$ represents the static component while $H_d(f)$ is the dynamic component which contains the vibration frequency $\omega$.

As discussed in this section, the vibration displacement $d$ correlates with the frequency $\omega$ of the swept sine chirp signal. Thus, both the amplitude and phase of received CSI measurements contain the vibration frequency information. For our system, we focus on analyzing the phase component. This is because the phase offset is more sensitive to subtle motion compare to amplitude. For example, the amplitude only provides granularity at centimeter-level while phase can sense micro-motion at millimeter-level.

### 4.4 CSI Calibration

Since the extracted CSI measurements are the sampled version of channel frequency response, such measurements incur significant distortions due to the hardware limitation of COTS WiFi NICs. Such distortions are primarily caused by clock unsynchronization at both the transmitter and receiver end. In order to achieve accurate motion sensing, we first adopt the error mitigation approach from previous work [27, 40] for CSI calibrations. In particular, we mitigate the sampling time/carrier frequency phase offsets by conducting conjugate multiplication using CSI measurements from two different antenna transmission pairs. Then, we select the pair where its CSI phase change yields higher power from the conjugate multiplication. Such selection strategy enables us to extract the CSI measurements contain richer information induced by vibration motions. We also apply a Butterworth high-pass filter with cut-off frequency at 100Hz to remove noises caused by motions at lower frequency range (e.g., human motion at various scales). Moreover, we utilize spectral subtraction techniques to further remove the stationary or slowly changing noises. It is done by subtracting the spectrum when there is no vibration motion from the spectrum with intensive vibration motion. Based on Equation (1), we assume that the output of conjugate multiplication is denoted as $\bar{H} = H_l H_s^*$, where $H_l$ is the CSI measurement on the $l$ th receiving antenna, $H_s^*$ is the conjugate of the CSI measurement on the $s^{th}$ receiving antenna and $l \neq s$. Hence, at a certain time, $\bar{H}$ contains 30 CSI subcarriers and can be written as follow:

$$\bar{H} = H_l H_s^* = [H_{l,1} H_{s,1}^* \quad H_{l,2} H_{s,2}^* \quad \ldots \quad H_{l,30} H_{s,30}^*]. \tag{10}$$

### 4.5 Resonance Frequency Extraction and Selection

In this section, we discuss how to extract and select appropriate resonance frequency from the CSI measurements obtained from the previous step.

**CSI Subcarriers Combination Using PCA.** After going through *Data Processing*, we obtain the denoised CSI measurements with 30 subcarriers which can be denoted as $\bar{H}$. Then we calculate the phase of $\bar{H}$ as $\bar{P} = [e^{j\Theta_1(\omega)} \; e^{j\Theta_2(\omega)} \; \ldots \; e^{j\Theta_{30}(\omega)}]$, where $e^{j\Theta_n(\omega)}$ ($n = 1, 2, \ldots, 30$) represents the corresponding phase of every element in Equation 10 and it is related to vibration frequency $\omega$. Thus, $\bar{P}$ is a matrix with dimensions $1 \times 30$. To capture the subtle vibration motion of a container, we adopt Principal Component Analysis (PCA) [48] by combining the 30 subcarriers information of $\bar{P}$. It is often used for dimensionality reduction of high-dimensional data. We select the first principle component that contains major and consistent phase variations caused by the vibration of the container. Assuming that at a certain time, the first principle component $P_{first}$ with dimensions $1 \times 1$ can be written as:

$$P_{first} = V\bar{P}, \tag{11}$$

where $V$ is the feature vector formed using the maximum eigenvector of the covariance matrix of $\bar{P}$.

**Frequency Domain Analysis.** In order to extract accurate resonance frequency information, we first need to conduct frequency analysis on the $P_{first}(t)$ which represents the first principle component of the CSI phase. Our system adopts the widely used frequency analysis method by applying the short-term Fourier transform (STFT) to the $P_{first}(t)$. After that, we obtain the spectrogram of the CSI phase component. For example, Figure 6(a) shows the spectrogram of a partially filled metal container with the first two resonance frequencies. Specifically, we use a Hamming window with a segment length of 2048 samples and 2048 sampling points to calculate STFT. Therefore, the frequency domain resolution is $2000/2048 \approx 0.98$Hz which provides sufficient resolution to capture the resonance frequency accurately. To further improve the time resolution, 2000 sampling points overlapped is set between segments. Figure 6(b) shows the first two resonance frequencies in the frequency responses of CSI phase change which represents the vibration frequency of the container in the corresponding spectrogram.

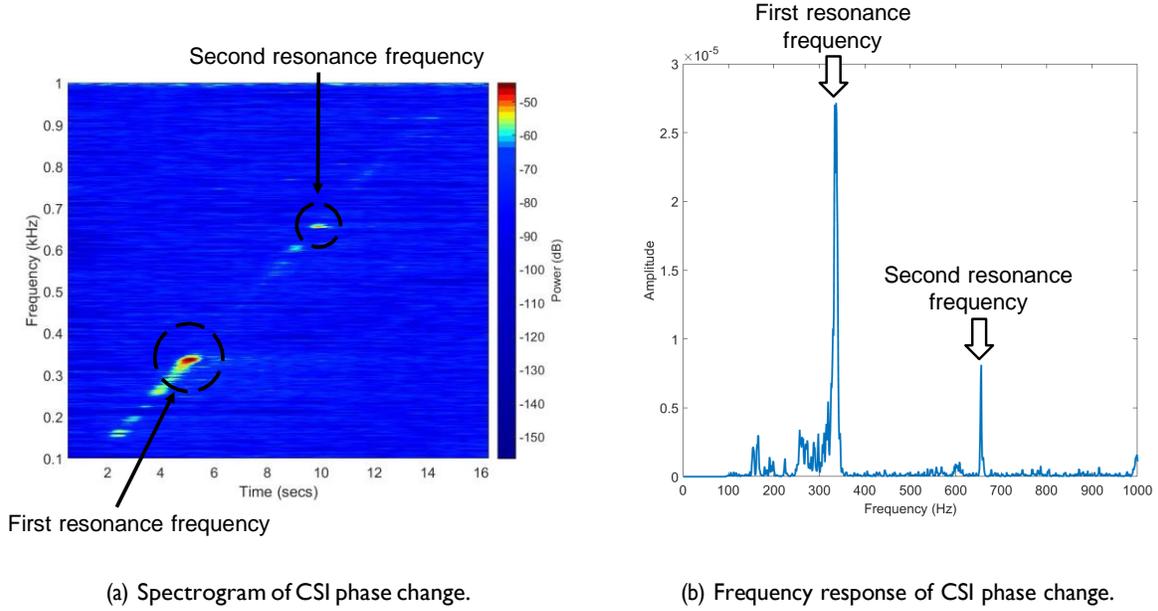

(a) Spectrogram of CSI phase change.  (b) Frequency response of CSI phase change.

Fig. 6. Spectrogram and frequency response of CSI phase change.

**Resonance Frequency Extraction and Selection.** After obtaining the spectrogram of the CSI phase, our system will extract the value of resonance frequency which is equivalent to the vibration frequency. As discussed before, the power of frequencies is unstable and it cannot reflect the characteristics of vibration frequency. As shown in Figure 6(b) we observe the frequency of CSI phase change at the selected liquid level presents an impulse pattern. This observation suggests that we can identify the first resonance frequency by extracting the first peak in frequency response. Since the power of the noise is trivial compared to the target frequency, thus we set the threshold as $Threshold = \frac{Maximum\ power\ of\ the\ frequency}{3}$. We then apply such a threshold to extract all the possible peaks.

In order to filter out the fake peaks, we apply a verification window with a length of 200Hz based on the observation that minimum distance between two neighboring peaks is usually larger than 200Hz. Previous work [44] indicated that the second resonance frequency is approximately 9/4 times first resonance frequency, the third resonance frequency is approximately 16/9 times second resonance frequency and so on. According to our experiments, the minimum frequency is about 170Hz. Thus, the minimum distance between two neighboring peaks is $170 \times 9/4 - 170 = 212.5$Hz. Then we select the first detected peak in frequency response as the first resonance frequency. In addition, we confirm the extracted frequency by comparing its value to multiple data

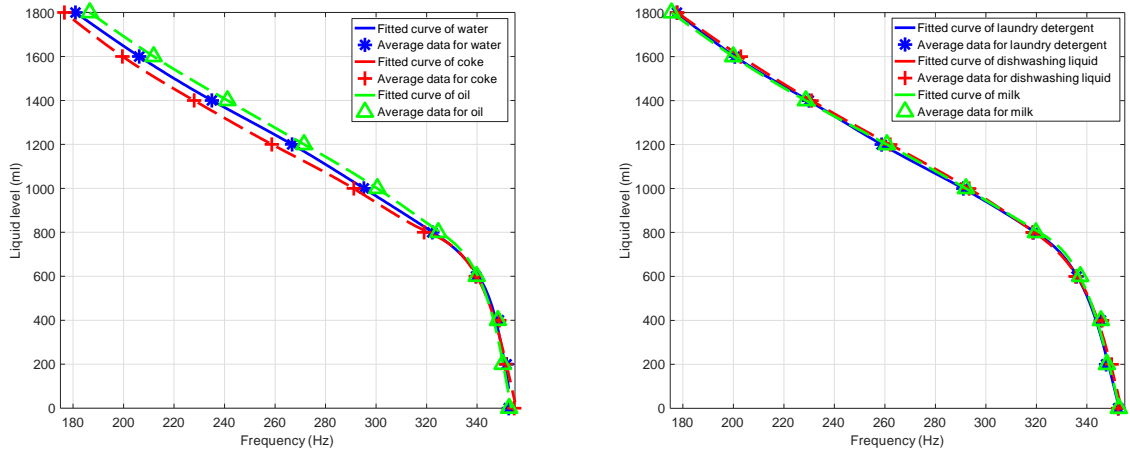

(a) Curve fitting for thin liquids: water, coke and vegetable oil.

(b) Curve fitting for thick liquids: laundry detergent, dishwashing liquid and milk.

Fig. 7. Piecewise polynomial curve fitting.

samples. Because our system will perform the frequency extraction procedure multiple times at a given liquid level and the results should be consistent.

### 4.6 Model Construction and Level Prediction

**Continuous Prediction Model.** For continuous prediction, we adopt curve fitting to construct the model. Figure 7 shows the average frequency values with respect to different liquid volumes for both thin and thick liquids in a metal container. In general, the liquid level and the resonance frequency are negatively correlated. This is due to the characteristics of the object resonance which the larger mass of the object is, the smaller the resonance frequency will be. However, such correlation can not be simply defined using a linear or polynomial function. As we can observe in Figure 7 (a) and (b), when the liquid level changes from high to medium (i.e., from 1800ml to 800ml), the resonance frequency change is relatively quick which can almost be perceived as a linear function. When the liquid level changes from medium to low (i.e., from 800ml to 0ml), the resonance frequency change is comparably slow which is similar to a polynomial function. Therefore, in order to achieve the best curve fitting, we adopt the piecewise polynomial interpolation. We can fit a low-degree polynomial also known as spline between each pair of consecutive data points. In other words, the fitted curve will be defined piecewise on the subintervals. Thus, we may have a different formula on each of the subintervals. To achieve a smoother fit, we use a cubic polynomial on each subinterval.

Figure 7 also shows the curve fitting results using piecewise polynomial. In particular, we have following frequency-level pairs: $(f_1, l_1), (f_2, l_2), \ldots, (f_i, l_i)$, where $i$ is the number of frequency-level pairs. Let $S(x)$ be the piecewise polynomial composed of $S_1(x), S_2(x), \ldots, S_{i-1}(x)$, where $S_h(x)$ denotes the spline on the $h^{\text{th}}$ subinterval $[x_h, x_{h+1}]$, for $h = 1, 2, \ldots, i-1$. If each $S_h(x)$ is a cubic polynomial, for $h = 1, 2, \ldots, i-1$ and $x_h < x < x_{h+1}$, we can write each cubic polynomial $S_h(x)$ in the form of:

$$S_h(x) = a_h + b_h(x - x_h) + c_h(x - x_h)^2 + d_h(x - x_h)^3, \qquad (12)$$

where $a_h$, $b_h$, $c_h$ and $d_h$ are four coefficients. Thus, there are $4(i - 1)$ unknowns to evaluate to solve all the coefficients. One condition requires the spline function to pass the first and last points of the interval, yielding $2(i-1)$ equations. Another condition requires that the first derivative is continuous at each interior point, yielding $(i - 2)$ equations. The third condition requires that the second derivative is continuous at each interior point, yielding $(i - 2)$ equations. To determine the last two equations, we use natural end conditions and clamped end conditions. Therefore, we can construct a continuous prediction model for our system. To achieve level prediction, we calculate the liquid volume using acquired resonance frequency based on a pre-built curve fitting model.

We found that the liquids with similar densities have similar resonance frequencies and liquid levels relationship. As shown in Figure 7, the curve fitting results from the same category of liquids (i.e., thin liquids or thick liquids) are similar. This is due to the fact that the density of the liquid is the only factor that affects the mass of the liquid, given the same amount of the liquids in the container. Thus, one model can be applied to different liquids that have similar densities with comparable performance. This enables us to use the model built from one liquid to different types of liquids, which presents tremendous training effort reduction for our system.

**Discrete Prediction Model.** For discrete prediction, we adopt a Support Vector Machine (SVM) based approach to build the model. In our system, we choose LIBSVM [7] toolbox to create a classification model. Specifically, the linear function kernel is used as we observed that the data is linearly separable. Moreover, resonance frequencies with labels are utilized as training data. We adopt a grid search method to find the optimal parameter of cost and gamma and leave other parameters as default.

Similar to continuous prediction models, discrete prediction models can also have the opportunity to reduce the training load. Moreover, it is possible that we can achieve an unsupervised discrete prediction model. The inherent relationship between the liquid level and the resonance frequency is that the higher the liquid level, the lower the resonance frequency. Based on this relationship, we may leverage cluster analysis (e.g. K-means) to achieve discrete liquid level prediction. Although such an unsupervised approach does not need labeling efforts, it requires a longer time to accumulate enough data to achieve high performance. Thus, this work, we only focused on SVM-based approach and leave the unsupervised approach in the future work.

## 5 PERFORMANCE EVALUATION

In this section, we will describe the experimental setup and present the performance of our system in sensing the liquid level of different types of liquids under various scenarios.

### 5.1 Experimental Setup

**Device and Network.** We conduct experiments with two laptops (i.e., Dell LATITUDE E5540). Both laptops run Ubuntu 10.04 LTS and are equipped with Intel 5300 WiFi NICs for extracting CSI measurements [19]. Specifically, we use one of the laptops as the transmitter and the other one as the receiver. The transmitter connects to one antenna, whereas the receiver connects to three antennas. The distance between the transmitter and the receiver is 1 meter if not specified. The target container is placed in the middle of the transmitter and receiver. For the experiments, a single 20MHz channel of 5GHz band was used and the package rate is set to 2000 pkts/s.

As shown in Figure 8(a), the transducer used in our system has the dimensions of 21.5mm × 14.5mm × 7.9mm. Its impedance is 8 Ohm with power handling of only 1 Watt. We connect the transducer to the control board to study the impact of the frequency band, sweep rate and vibration intensity on the performance of LiquidSense and to determine the optimal parameters. Once these parameters have been optimized, the control board is not needed in the real system deployment. Such a small and low-cost transducer can be easily mounted on the surface of any container (e.g. a metal container) as shown in Figure 8(b). During the experiments, our system utilizes the chirp signal as input and keep the intensity as constant. To fully evaluate LiquidSense's performance

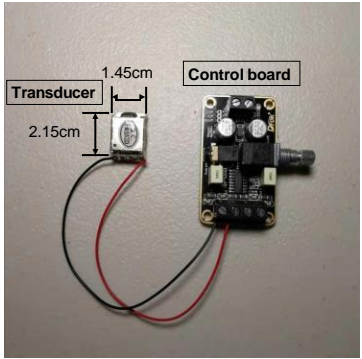

(a) The low-cost small transducer and the control board.

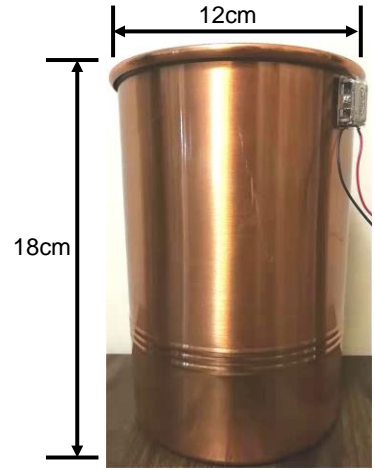

(b) The transducer is mounted on the surface of a metal container.

Fig. 8. Transducer and container.

Table 1. Three different containers used in our experiments.

| Material | Height (cm) | Diameter (cm) | Capacity (ml) |
|---|---|---|---|
| Metal | 17 | 7.4 | 720 |
| Metal | 18 | 12 | 1800 |
| Glass | 18.5 | 13 | 2160 |
| Ceramic | 15 | 18.5 | 3600 |

Table 2. Six different types of liquids used in our experiments.

| Liquid | Density (g/cm$^3$) |
|---|---|
| Vegetable oil | 0.93 |
| Water | 1.00 |
| Coke | 1.01 |
| Milk | 1.03 |
| Laundry detergent | 1.04 |
| Dishwashing liquid | 1.06 |

over different containers, we choose four commonly used containers for liquids in daily life as shown in Table 1. They are made up of the following materials: metal, glass, and ceramic. The capacities of these containers are 720ml, 1800ml, 2160ml and 3600ml, respectively.

**Environments and Types of Liquids.** The experiments are conducted in a typical home environment with six different types of liquids that are commonly found in daily life. The densities of liquids are listed in Table 2. Those liquids fall into two categories based on their density: thin liquids and thick liquids. The thin liquids have lower density including water, coke and vegetable oil, while the thick liquids have higher density including milk, dishwashing liquid, and laundry detergent. The aforementioned liquids cover both edible and non-edible categories for everyday use. Figure 9 illustrates a typical experimental setup of our system. If not specified, the typical distance between the transmitter and the receiver is 100cm and the testing container is placed in the middle of the transmission pair. We mount the transducer on both the upper part and the lower part of the surface of the container to study the impact of the transducer's positions. Figure 8(b) shows one example that the transducer is mounted on the upper part of the surface of the container. If not specified, the reported result is under the placement of the upper part of the container. We study how the distance between the WiFi devices affect the performance at distances of 50cm, 150cm, 250cm, 350cm and 450cm, respectively. We also vary the angle between the transmitter and the receiver including 60, 120 and 180 degrees. In the NLOS scenario, the transmitter and the container are separated by a wall.

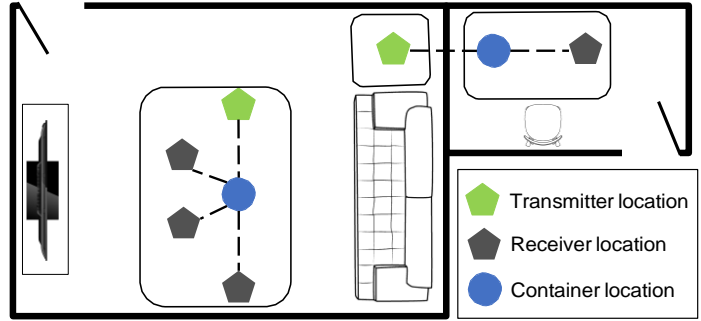

Fig. 9. Illustration of experimental setup.

### 5.2 Metrics

We use the following metrics to evaluate the performance of our system.

**Continuous Prediction Metrics.** For continuous prediction, we use the error rate as a metric which is the ratio of the difference between the predicted value and the ground truth to the total capacity of the container. The ground truth of a certain liquid level can be measured by a graduated cylinder. Thus, the error rate can be represented as follow:

$$Error\ Rate = \frac{|Predicted\ Value - Ground\ Truth|}{Total\ Capacity}.$$

Therefore, the system accuracy can be written as $Accuracy = 1 - Error\ Rate$.

**Discrete Prediction Metrics.** For discrete prediction, we use confusion matrix as well as F-score to evaluate the performance. In a confusion matrix, each row represents the ground truth of a certain liquid level and each column shows the predicted liquid level that was classified by our system. Each cell in the matrix corresponds to the fraction of liquid level in the row that was classified as the liquid level in the column. F-score is the weighted average of Precision and Recall.

### 5.3 Data Collection and Training

For the data collection procedure, the transducer is set to output both positive excitation from 0Hz to 1000Hz and negative excitation from 1000Hz to 0Hz chirp signals with the sweep rate of 66.67Hz per second. Here, we denote one sweep including one positive excitation and one negative excitation, as one sweep will generate only one resonance frequency. Each WiFi probing session is set to 17s in order to fully capture the vibration of the container at the resonance frequency.

In our experiments, we use three different containers (i.e., metal, glass and ceramic) and six types of liquids, which can be further divided into two categories based on density as discussed before. We divide the capacity of each container into 9 equal parts and thus we obtain 10 levels including the empty level. Then we perform

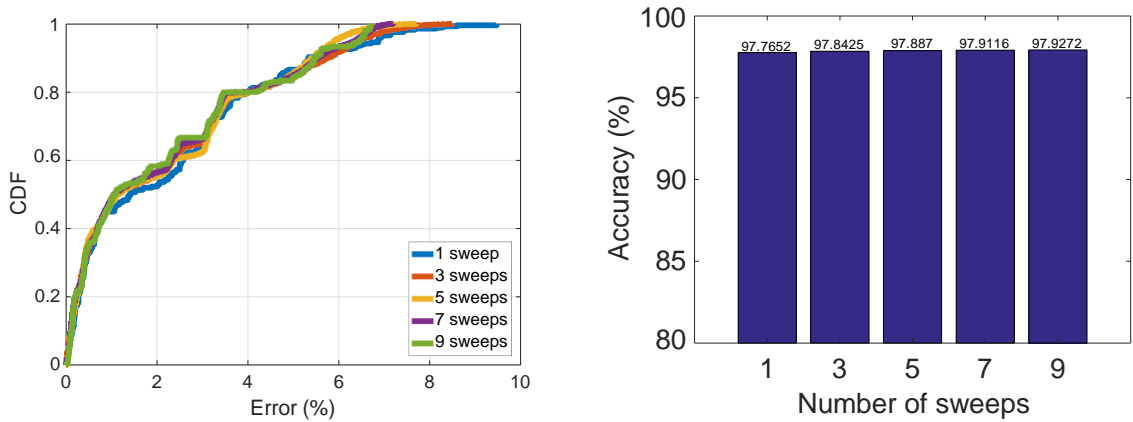

(a) Overall CDF of errors of continuous liquid level under different numbers of sweeps.

(b) Overall accuracy of continuous liquid level prediction under different numbers of sweeps.

Fig. 10. Overall continuous prediction performance of LiquidSense.

10 sweeps at each level for each container filled with one type of liquid. Thus, we obtain 100 samples for one container with one type of liquid. For the metal container, we conduct an experiment with six types of liquids (600 samples in total). For the glass and the ceramic container, we choose one type of liquid from each category including water and dishwashing liquid (200 samples each container and 400 samples in total). We also use the metal container and water as the liquid to conduct experiments under different distances, angles and NLOS scenarios (500 samples in total).

For continuous prediction, the training dataset contains the data for $1^{st}$, $3^{rd}$, $5^{th}$, $7^{th}$ and $10^{th}$ levels, whereas the testing dataset includes the data for $2^{nd}$, $4^{th}$, $6^{th}$, $8^{th}$ and $9^{th}$ levels, which is non-overlapping with the liquid levels used in training. Note that the data used in every evaluation are collected on different days. We use 10 samples from each level and calculate the average resonance frequency for training. This results in 50 samples for one container with one type of liquid (500 samples in total) as continues prediction model training. For discrete prediction, we label the liquid levels from low to high with 1 to 10. We randomly select half of the data (50 samples) on each liquid level as the training dataset, and then use the remaining half of the data (50 samples) as the testing dataset.

### 5.4 Overall Performance

Figure 10(a) shows the overall error CDF of continuous prediction for a metal container with six different types of liquids under different numbers of sweeps. We can observe that all the errors are ranging from 0.1% to 10% and 90% of the errors are less than 6%. This shows our system is highly accurate in sensing the liquid level in a continuous manner. Moreover, the overall accuracy is great than 97% as shown in Figure 10(b). Although it is possible to further increase the accuracy of our system using additional rounds of sweeping, our system can already achieve high accuracy with only one round of sweep. Thus, we only reported the results of one sweep in the rest of this paper. Figure 11(a) shows the confusion matrix of discrete liquid level prediction of a metal container with six types of liquids. The overall precision, recall and F-score are 0.968, 0.967 and 0.968 respectively as shown in Figure 11(b). It is worth noticing that the majority of the errors occur when contain has a lower

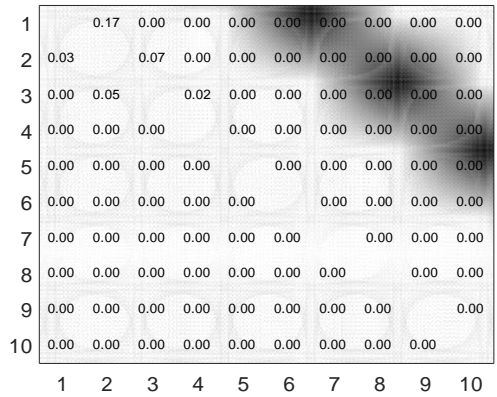
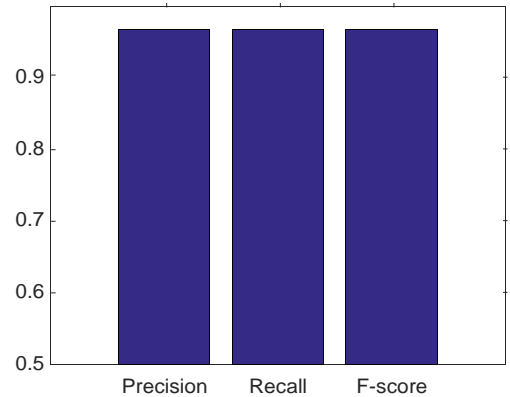

(a) Overall confusion matrix of discrete liquid level prediction.

(b) Overall Precision, Recall and F-score of discrete liquid level prediction.

Fig. 11. Overall discrete prediction performance of LiquidSense.

liquid level. This is due to that with less liquid, the resonance frequency changes at a slower pace. On the other hand, when contain has a higher liquid level, our system can achieve better performance.

## 5.5 Impact of Different Types of Liquids

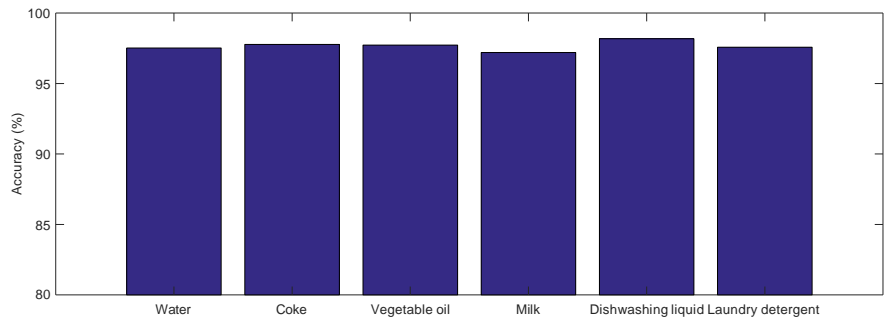

Fig. 12. Continuous prediction accuracy of a metal container with six types of liquids.

We next evaluate the impact of liquid type on the performance of our system. As discussed before, six different types of liquids are divided into two categories based on their densities: thin liquids and thick liquids. Water, coke and vegetable oil are considered thin liquids which have relatively low density. Meanwhile, milk, dishwashing liquid, and laundry detergent have a relative high density which are in the thick liquid category. The results are shown in both Figure 12 for continues prediction and Table 3 for discrete prediction. We can observe from Figure 12 that the average accuracy of continuous prediction for different types of liquids is over 97.3% and there are no significant differences between thin and thick liquids. Table 3 shows the results of discrete liquid level prediction. The F-score of our system is above 0.93 for all the target liquids. This demonstrates our system can achieve high accuracy under both continues and discrete prediction model while applying to a wide variety of liquids.

Table 3. Discrete prediction performance of a metal container with six types of liquids

| Liquid | Precision | Recall | F-score |
| --- | --- | --- | --- |
| Water | 0.9417 | 0.9400 | 0.9318 |
| Coke | 0.9769 | 0.9700 | 0.9737 |
| Vegetable oil | 0.9714 | 0.9600 | 0.9657 |
| Milk | 0.9909 | 0.9900 | 0.9905 |
| Dishwashing liquid | 0.9833 | 0.9800 | 0.9817 |
| Laundry detergent | 0.9714 | 0.9600 | 0.9657 |

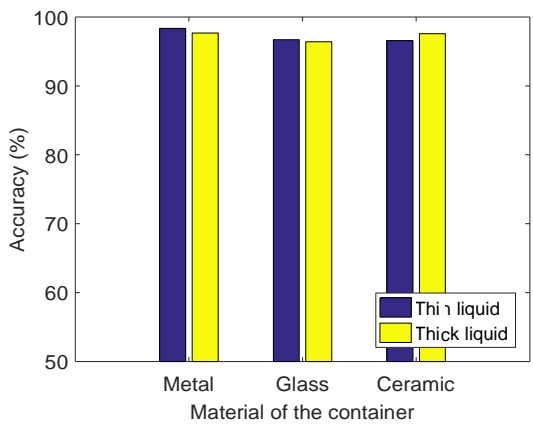

Fig. 13. Continuous prediction accuracy of different materials of containers.

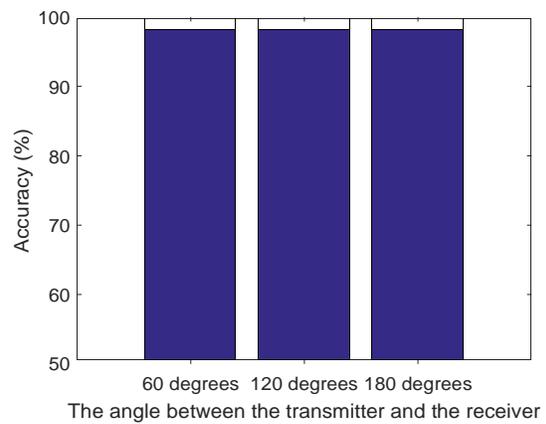

Fig. 14. Continuous prediction accuracy under different angles.

## 5.6 Impact of Different Materials of Containers

We then evaluate the impact of container material by choosing three containers made of commonly used materials (i.e., metal, glass and ceramic). Both categories of liquids are considered: thin liquid (water) and thick liquid (dishwashing liquid). As shown in Figure 13, our system achieves over 96.3% accuracy for three containers made of different materials while filled with thin and thick liquids respectively. Thus, our continuous prediction model works well for containers with different materials as well as liquids with different densities. Table 4 shows the results of discrete prediction under different containers and liquids scenarios. We can observe that the F-score is over 0.92. Since the ceramic container has the largest volume comparing to the other two containers, equal proportions of liquid will result in greater resonance frequency changes. Therefore, it is easier to distinguish different liquid levels for such a container using discrete prediction model. In a nutshell, LiquidSense can be applied to a variety of containers of different materials.

Table 4. Discrete prediction performance of different materials of containers.

| Material/Density of Liquid | Precision | Recall | F-score |
|---|---|---|---|
| Metal/Thin | 0.9417 | 0.9400 | 0.9318 |
| Metal/Thick | 0.9833 | 0.9800 | 0.9817 |
| Glass/Thin | 0.9294 | 0.9300 | 0.9297 |
| Glass/Thick | 0.9526 | 0.9500 | 0.9513 |
| Ceramic/Thin | 1.0000 | 1.0000 | 1.0000 |
| Ceramic/Thick | 0.9909 | 0.9900 | 0.9905 |

Table 5. Discrete prediction performance under different angles.

| Angle | Precision | Recall | F-score |
|---|---|---|---|
| 60 degrees | 0.9769 | 0.9700 | 0.9734 |
| 120 degrees | 0.9833 | 0.9800 | 0.9817 |
| 180 degrees | 0.9909 | 0.9900 | 0.9905 |

### 5.7 Impact of Different Degrees

Because the relative position between the transmission pair and the container is subject to change from time to time, we further evaluate the impact of different angles of the container with respect to transmission pair. The distance between the transmitter and the container is equal to the distance between the receiver and the container which is 25cm. The angles refer to the angle between the transmitter and the receiver which are 60 degrees, 120 degrees, and 180 degrees. We only use the metal container filled with water for this study. Overall, we observe that our system could achieve similar performance even under different angles scenarios. In particular, the accuracy of continuous prediction is over 97.6% as shown in Figure 14. Meanwhile, the F-score of discrete prediction is greater than 0.95 as shown in Table 5. Since the vibration of the container at resonance frequency occurs in all the inner and outer surface area of the container simultaneously. Even when the angle between the container and transmission pair changes, our system can still capture the vibration of the container using commodity WiFi.

### 5.8 Impact of Different Distances

Next, we study the impact of distance between transmitter and receiver on the performance of our system. The experimental distances are 50cm, 150cm, 250cm, 350cm and 450cm, respectively. Figure 15(a) shows that our system has an average prediction accuracy of 98.2%, 96.1%, 93.3%, 88.5% and 83.1% at a distance of 50cm, 150cm, 250cm, 350cm and 450cm, respectively. As shown in Figure 15(b), the F-scores of 50cm, 150cm, 250cm, 350cm and 450cm are 0.9905, 0.8953, 0.82, 0.7561 and 0.7073, respectively. We observe that our system achieves an accuracy of 93.3% and the F-score of 0.82 at 250cm, which is a typical distance between a transmitter and a receiver in a single room. By reducing the distance to 50cm, we can improve the system accuracy to 98% and the F-score to 0.99. This is because a shorter transmission distance leads to higher received signal strength and a higher signal to noise ratio. Indeed, the system performance decreases at a longer distance (e.g., 450cm) in a cross-room setting. However, we imagine that there will be multiple smart and IoT devices with WiFi interface available in a typical smart home environment. We could then choose one transmitter-receiver pair among these multiple devices to reduce the sensing distance and to improve the system performance. Thus, the high density of smart and IoT devices in smart homes provides us the opportunity to place the container in the effective range (i.e., less than 250cm) of LiquidSense. We could also move or deploy smart and IoT devices close to the container so as to have a satisfactory system performance.

### 5.9 Impact of Non-Line-of-Sight

In this section, we study the impact of non-line-of-sight (NLOS) by placing the WiFi transmitter and the WiFi receiver in two adjoining rooms with a closed door. The container is placed on the receiver side. We study

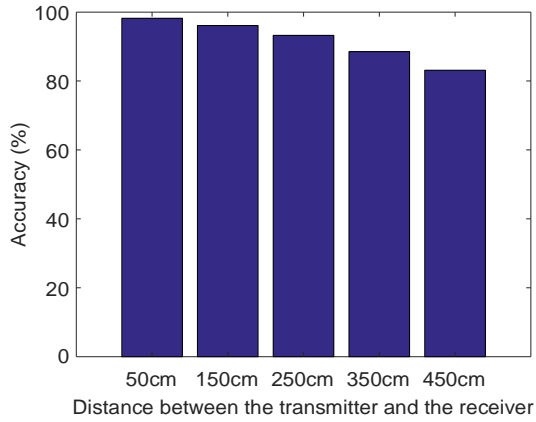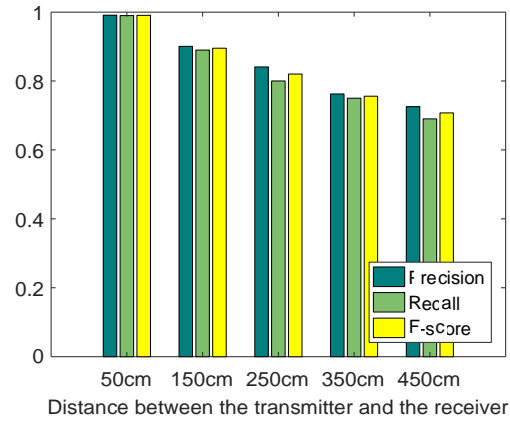

(a) Accuracy of continuous prediction under different distances.

(b) Precision, Recall and F-score of discrete prediction under different distances.

Fig. 15. System performance under different distances.

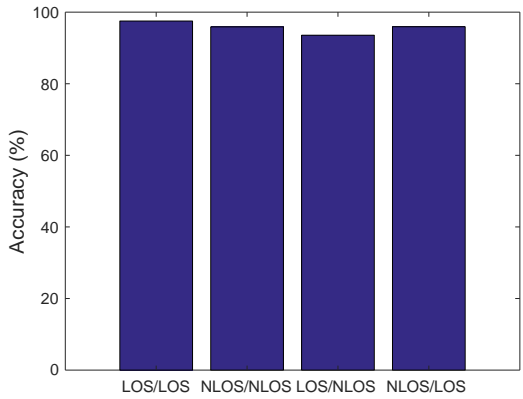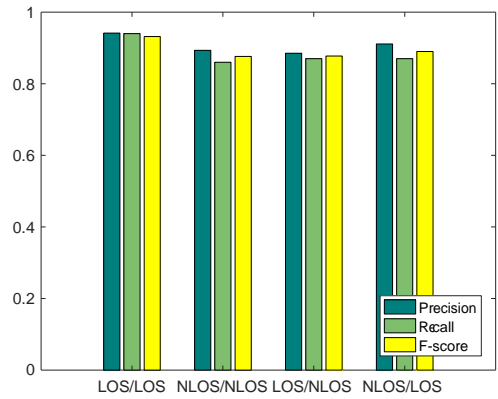

(a) Accuracy of continuous prediction under LOS and NLOS.

(b) Precision, Recall and F-score of discrete prediction under LOS and NLOS.

Fig. 16. System performance under both LOS and NLOS scenarios.

the performance of our system by using the training data collect under LOS scenarios to test the data collected under NLOS scenarios (i.e., LOS/NLOS in Figure 16). We also show the results by using the training data collected under NLOS scenarios to test the data under LOS scenarios (i.e., NLOS/LOS in Figure 16). For comparison, we also present the results when both the training and testing data are collected under either the NLOS or LOS scenarios (i.e., LOS/LOS and NLOS/NLOS in Figure 16). As shown in Figure 16(a), the system accuracy exceeds 93.5% in all four evaluation scenarios in continuous prediction. Meanwhile, the F-scores in these four evaluation

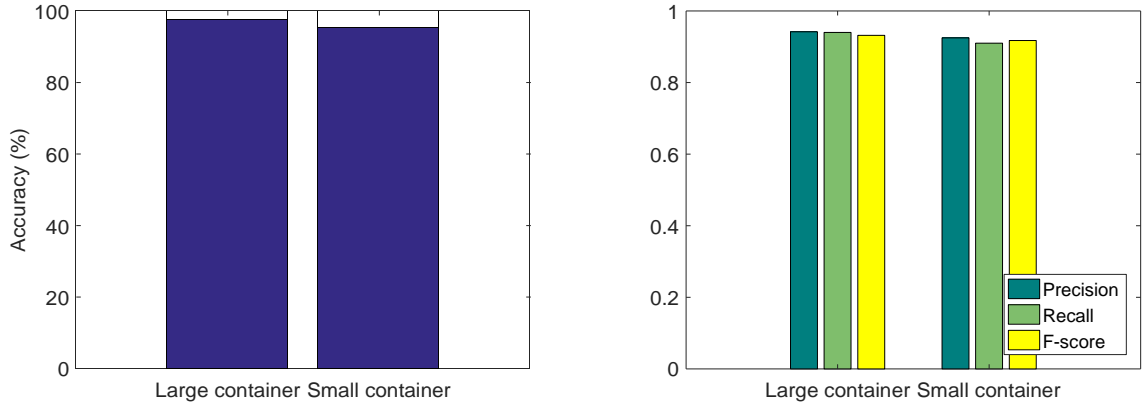

(a) Accuracy of continuous prediction with containers of different sizes.

(b) Precision, Recall and F-score of discrete prediction with containers of different sizes.

Fig. 17. System performance of a larger container and a small container.

scenarios are around 0.9 in discrete prediction. The above results indicate that without distinguish between LOS and NLOS in LiquidSense, our system still achieves very accurate liquid level detection. This is because our system is able to accurately detect the resonance frequency under both LOS or NLOS and the resonance frequency of the container remains the same when the environment changes. Moreover, we observe that using the training and testing data under the same LOS or NLOS scenarios could marginally improve the performance. We thus could leverage existing work in distinguish NLOS and LOS, such as PhaseU [64] and the work presented in reference [65], to fine-tune the prediction results.

### 5.10 Impact of Container Size

We then discuss the impact of container size by evaluating two metal containers with different sizes. The large container has a height of 18cm, a diameter of 12cm, and a capacity of 1800ml. The small container has a height of 17cm, a diameter of 7.4cm, and a capacity of 720ml, which is about one-third of the capacity of the large container. For continuous prediction, Figure 17(a) shows that LiquidSense achieves an accuracy of 97.6% when using a large container and an accuracy of 95.3% when using a small container. As shown in Figure 17(b), the F-scores of two containers are 0.9318 and 0.9172, respectively. The results show that our system works well with a small container, although the performance can be slightly improved by using a large container. The is mainly because a larger container will affect more multipath of WiFi signals, which results in slightly better accuracy.

### 5.11 Impact of Transducer Position

In all of the evaluations described so far, we mounted the transducer on the upper part of the container. The users may install the transducer at different positions and the position of the transducer may affect the system performance. We thus study how the position of the transducer affects the performance of our system. We collect data with the transducer mounted on both the upper part and lower part of the surface of the container. Figure 18(a) shows that the upper part scenario has an accuracy of 95.3% and the lower part scenario has an accuracy of 91.5%, both in continuous prediction. Figure 18(b) shows the results of discreet prediction, in which

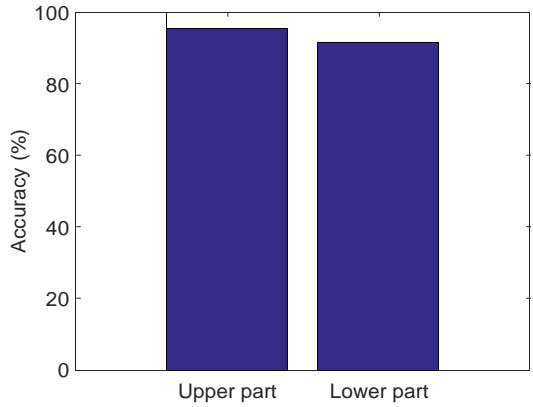
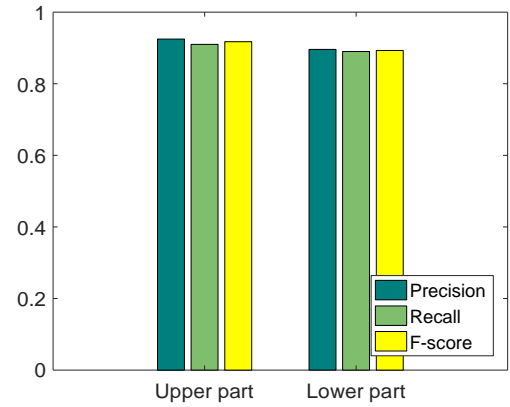

(a) Accuracy of continuous prediction under different positions of the transducer.

(b) Precision, Recall and F-score of discrete prediction under different positions of the transducer.

Fig. 18. System performance under different positions of the transducer.

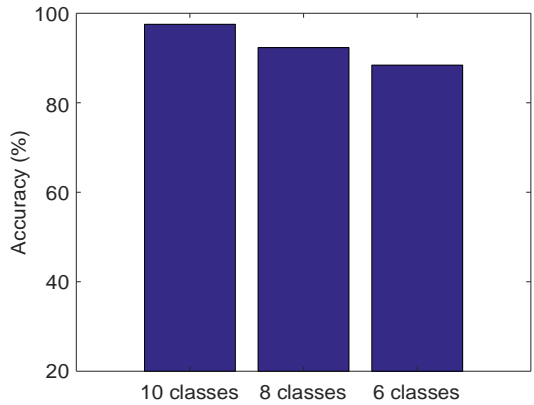
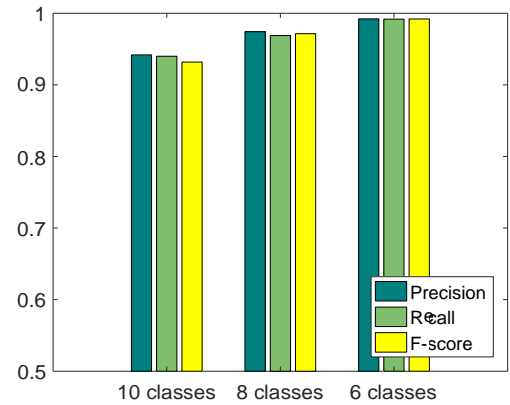

(a) Accuracy continuous prediction of different classes.

(b) Precision, Recall and F-score of discrete prediction of different classes.

Fig. 19. System performance under different number of classes.

the upper part scenario has the F-score of 0.9172 and the lower part scenario has the F-score of 0.8928. The results show that mounting the transducer on the upper part of the container achieves a better performance as the higher the position where the transducer is mounted, the easier the container vibrates at large amplitude. Thus, we recommend installing the transducer on the upper part of the container to have a better system accuracy.

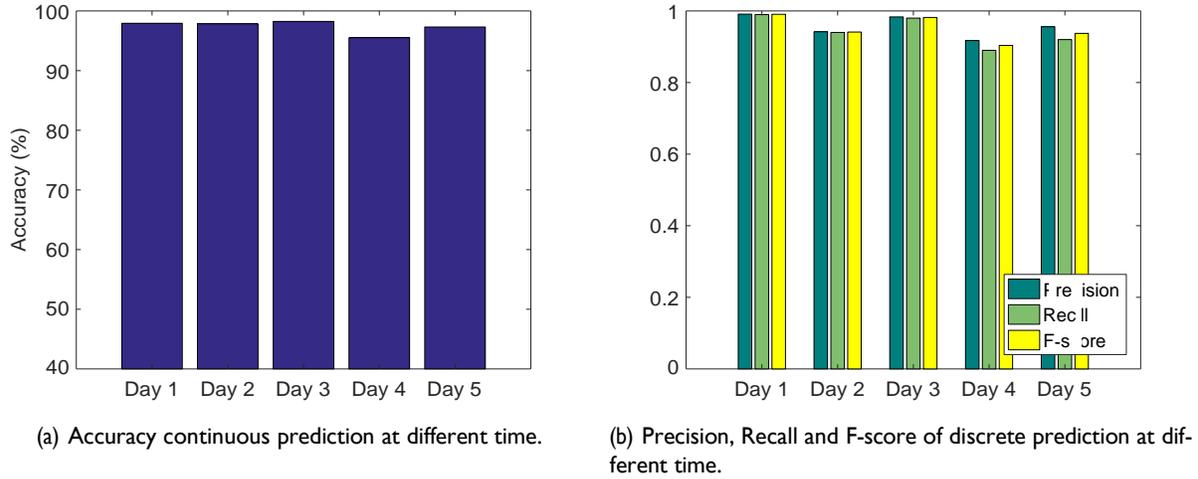

(a) Accuracy continuous prediction at different time.

(b) Precision, Recall and F-score of discrete prediction at different time.

Fig. 20. System performance at different time.

### 5.12 Impact of Different Number of Classes

In this section, we investigate the impact of using a different number of classes as training data for liquid level prediction. Technically, using more classes to fit the curve will produce a better or a more smooth relationship between the resonance frequency and the liquid level in continuous prediction, which could improve the accuracy of liquid level prediction. However, increasing the number of classes in the training data makes it more difficult for discrete classification, which will degrade the accuracy in discrete classification. We study such impacts by using the training data that includes 6 classes, 8 classes and 10 classes for both discrete and continuous prediction, respectively. Figure 19(a) shows that our system has an accuracy of 97.5%, 92.3% and 88.4% in continuous prediction under 10 classes, 8 classes and 6 classes, respectively. Meanwhile, Figure 19(b) shows that LiquidSense achieves the F-score of 0.9318, 0.9715 and 0.9919 in discrete prediction under 10 classes, 8 classes and 6 classes, respectively. These results match the technical analysis: increasing the training data classes improves the accuracy of continuous prediction. Still, our system produces accurate sensing results under both continuous and discrete prediction with a reasonable number of training classes.

### 5.13 Robustness Against Time and Environmental Changes

In wireless sensing, system performance could vary over time since the environment changes over time. Thus, we study the robustness of our system against time and environmental changes. We use the dataset of Day 1 as the training set and use the datasets of Day 2, Day 3, Day 4 and Day 5 as test sets. The environment also changed during these 5 days (e.g., the angle between WiFi devices, the distance between WiFi devices, positions of furniture and the movement of the human). Figure 20(a) shows that the continuous prediction accuracy changes over time and it is better than 95% in all cases. Similarity, Figure 20(b) also shows that our system has the F-score over 0.9034. The results show that our system is robust to time and environmental changes. The reason is that the resonance frequency is not sensitive to changes in time and environment, and the environmental changes only induce low frequencies signal change in the frequency domain, which can be easily filtered out by the high-pass filter in our system.

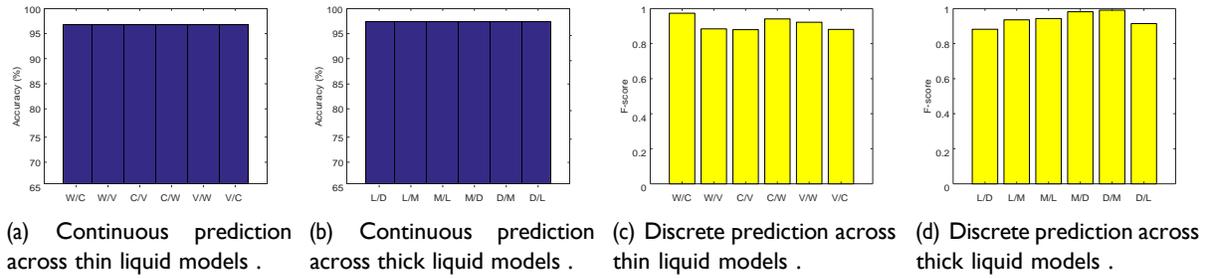

Fig. 21. System performance of different continuous and discrete prediction models.

### 5.14 Robustness of LiquidSense across Different Prediction Models

Last, we study the robustness of our system by using only liquid density-based prediction models (i.e., thin and thick) and testing across different types of liquids within various containers. For simplicity, we denote each type of liquid using its initial cap (e.g., W as water, C as coke and so on). We use the pre-built model of a certain liquid to perform prediction on a different type of liquid. For example, W/C represents that we use water's model to predict coke's level. Figure 21(a) and (b) show that the accuracy of our system is greater than 95.5% in continuous prediction across thin liquid models and greater than 96.9% in continuous prediction across thick liquid models. Figure 21(c) and (d) show that the F-score of discrete predictions across thin and thick liquid models is over 0.88. Thus, the prediction model based on liquid density is robust across various types of liquids as well as containers.

## 6 DISCUSSION

**Sensing Multiple Containers.** Currently, we only evaluate our system with a single container present in the environment. However, there could be multiple containers in a kitchen or in a restaurant. To deal with the multiple containers, we could extend our system with "multiple access" compatibility including both Time Division Multiple Access (TDMA) and Frequency Division Multiple Access (FDMA). In particular, TDMA allows us to sense multiple containers in different time slots. We can easily separate different containers in the time domain when performance resonance frequency sensing. Moreover, we could also assign different sensing frequency bands to different containers, which is an analog of frequency division multiple access. To accommodate a large number of containers, for example, in a restaurant, we could combine both the TDMA and FDMA to improve the discrimination power. We thus would like to study the effectiveness and the efficiency of both TDMA and FDMA methods for multiple containers sensing in our future work.

**Energy Consumption.** In our work, vibration generation relies on a battery, which has a limited capacity. To improve the power efficiency, our system performs resonance frequency sensing periodically each day. For example, sensing 3 to 5 times each day should be sufficient for daily liquid level monitoring. For each sensing, the duration of the vibration is less than 15s in our system. Therefore, a commercial off-the-shelf button cell battery could last for a few months when using a transducer with power handling of 1 Watt used in our experiments. To further reduce the energy consumption, we could optimize the sensing frequency band of the chirp signal to reduce the sensing duration. For example, we could reduce the frequency band to 250Hz, which covers all the resonance frequencies of the containers in our experiments. The energy consumption will be further reduced to one-quarter of the energy consumption of the chirp (with a frequency band of 1000Hz) that currently used in our experimental evaluation.

**Unsupervised Learning Method.** Our current discrete prediction method relies on SVM, which is a supervised learning method that requires labeled training data. However, labeling the data could be an expensive process in terms of time, labor and human expertise. We could reduce such human efforts by leveraging the inherent relationship between the liquid level and resonance frequency. That is the higher the liquid level, the lower the resonance frequency. Based on this fact, we may leverage cluster analysis (e.g. K-means) to achieve liquid level prediction in an unsupervised manner (i.e., without labeled training data). In particular, we could first collect the resonance frequencies for several usage cycles (i.e., from full to empty). Then, we can cluster these resonance frequencies into several groups, each representing a similar level of the liquid. Based on the inherent relationship between the liquid level and resonance frequency, we can infer the coarse-grained liquid level information and generate a resonance frequency and liquid level profile. After that, by searching the minimum distance of a new resonance frequency to each cluster, we can calculate the predicted liquid level. Of course, such a method is a trade-off between the prediction accuracy and the training effort, and will be investigated in our future work.

**Future Applications.** LiquidSense could enable a large number of useful applications in smart home and IoT environments in the further. It can provide visibility of existing household liquid quantities. Such knowledge can help users decide whether they need to replenish a product or refill patients' medication bottles. It can also provide knowledge about how much liquid in the container has been consumed. Thus, users can calculate calories ingested by themselves and their families. Then it is possible to provide the users with a suitable diet plan. Moreover, we can combine LiquidSense with other existing home or factory sensing methods. For example, motion sensors could be used to detect the motion of the container and trigger the container level sensing once the container has been touched. Thus, the system could report the liquid level change immediately and also save energy.

## 7 CONCLUSION

This paper presents LiquidSense, which is a low-cost, high accuracy, widely applicable liquid level sensing system. The proposed system leverages only one transducer and existing commodity WiFi devices to achieve liquid level sensing, which can be easily integrated with a smart home environment. The insight is that the transducer makes a container resonant and WiFi signals can capture the resonance frequency, which is associated with the liquid level in the container. By analyzing the frequency of CSI phase change of WiFi signals, we can extract resonance frequencies as features for liquid level. LiquidSense achieves both continuous and discrete liquid level prediction by using curve fitting and SVM classifier, respectively. Extensive experiments under different types of liquids and containers of different materials demonstrate that LiquidSense is effective in predicting a number of liquid levels and that it can achieve over 97 % accuracy in continuous prediction as well as an F-score of 0.968 in discrete prediction. Moreover, our system can work well under different angles between the transmitter and the receiver. In addition, we show that our system can provide accurate liquid level prediction under a relatively long distance or NLOS scenario.

## ACKNOWLEDGMENTS

We thank the anonymous reviewers for their insightful feedback. This work was partially supported by the NSF Grants CNS-1910519, CNS-1514238, and DGE-1565215.